\documentclass[10pt,conference]{IEEEtran}\IEEEoverridecommandlockouts
\usepackage{epsfig,graphicx,subfigure,psfrag,amsmath,cases}
\usepackage{latexsym,amssymb,amsmath,epsfig,subfigure,algorithm}
\usepackage{algorithmic}
\usepackage{color}
\usepackage{url}
\usepackage{scrtime}

\author{\authorblockN{Shiyang Leng, Derrick Wing Kwan Ng, and Robert Schober\authorrefmark{1}\thanks{\authorrefmark{1}The author is also with the University of British Columbia. This work was supported in part by the AvH Professorship Program of the Alexander von Humboldt Foundation.}}
Friedrich-Alexander-University Erlangen-N\"urnberg (FAU), Germany\\
Email: vicky.s.leng@studium.fau.de, kwan@lnt.de,  schober@lnt.de\vspace*{-3mm}
}

\title{Power Efficient and Secure Multiuser Communication Systems with Wireless Information and Power Transfer\vspace*{-2mm}}

\date{\thistime,\,\today}

\newtheorem{Thm}{Theorem}
\newtheorem{Lem}{Lemma}

\newtheorem{proposition}{Proposition}

\DeclareMathOperator{\Tr}{\mathrm{Tr}}
\DeclareMathOperator{\Rank}{\mathrm{Rank}}
\DeclareMathOperator{\nullspace}{\mathrm{Null}}

\newtheorem{Remark}{Remark}

\DeclareMathOperator{\maxo}{maximize}
\DeclareMathOperator{\mino}{minimize}

\newcommand{\abs}[1]{\lvert#1\rvert}
\newcommand{\norm}[1]{\lVert#1\rVert}

\begin{document}

\maketitle

\begin{abstract}
In this paper, we study   resource allocation algorithm design for power efficient secure communication with simultaneous wireless information and power transfer (WIPT) in multiuser communication systems.  In particular, we focus on  power splitting receivers which are able to harvest energy and decode information from the received signals.  The considered problem is modeled as an optimization problem which
takes into account  a minimum required signal-to-interference-plus-noise ratio (SINR) at multiple  desired receivers, a maximum tolerable data rate at multiple multi-antenna potential eavesdroppers,  and a minimum required power delivered to the receivers. The proposed problem formulation facilitates  the dual use of artificial noise in  providing  efficient
energy transfer and guaranteeing secure communication. We aim at minimizing the total transmit power by jointly optimizing transmit beamforming vectors, power splitting ratios at the desired receivers, and the covariance of the artificial noise.
The resulting non-convex optimization problem is transformed into a semidefinite programming (SDP) and solved by SDP relaxation. We show that the adopted SDP relaxation is tight and achieves the global optimum of the original problem. Simulation results illustrate the significant power saving obtained by the proposed optimal algorithm compared to suboptimal baseline schemes.
\end{abstract}

\renewcommand{\baselinestretch}{0.92}
\large\normalsize

\section{Introduction} \label{sect1}

The explosive growth of  high speed wireless communication has heightened the energy demand of communication networks. Handheld  mobile communication devices are
often powered by batteries with  limited energy storage capacity
which remains a bottleneck in prolonging the lifetime
of networks. As a result, the integration of energy harvesting capabilities into communication terminals
is considered as a promising solution for providing  self-sustainability to energy constrained wireless devices and thus has drawn significant interest recently.  Apart from conventional energy harvesting methods such as wind and solar, an emerging technology, wireless power transfer,  has been proposed to scavenge energy from the ambient radio frequency (RF)  signals \cite{CN:Shannon_meets_tesla}--\nocite{JR:WIP_receiver,JR:Rui_zhang_power_splitting,CN:Bruno_WIPT,JR:Kai_bin,JR:Sean_WIPT_analysis,
CN:WCNC_WIPT,JR:WIPT_fullpaper}\cite{JR:WIPT_fullpaper}. In particular, wireless power transfer serves the dual purpose of simultaneous wireless information and power transfer (WIPT). In \cite{CN:Shannon_meets_tesla}, the fundamental trade-off between channel capacity and harvested energy was studied. In \cite{JR:WIP_receiver}, a practical power splitting receiver was proposed to realize concurrent information decoding and energy harvesting for single user single antenna systems. This work was then extended to multiuser systems with multiple transmit antennas in \cite{JR:Rui_zhang_power_splitting}.  In \cite{CN:Bruno_WIPT} and \cite{JR:Kai_bin}, different transmission strategies were proposed to enable efficient WIPT.  In \cite{JR:Sean_WIPT_analysis}, the performance of WIPT  systems  was analyzed for different relaying protocols. In  \cite{CN:WCNC_WIPT} and \cite{JR:WIPT_fullpaper},  the energy efficiency of  multi-carrier systems  with  simultaneous WIPT was studied for different system configurations. In particular,  it was shown in   \cite{JR:WIPT_fullpaper} that the  energy efficiency of a communication system can be improved by integrating an energy harvester into a conventional information receiver. The results in \cite{CN:Shannon_meets_tesla}--\cite{JR:WIPT_fullpaper} suggest that  increasing  the transmit power of information signals facilitates both information decoding and energy harvesting at the receivers. However, the increased signal powers for WIPT makes the information signals  more vulnerable to eavesdropping due to a higher potential for information leakage. Thus, communication security in WIPT systems is  a critical issue.

Traditionally, cryptographic encryption technologies enable communication security in the application layer. However, there are some well-known drawbacks of  cryptographic encryption   such as high computational complexity and the required secure key distribution. As an alternative,  physical (PHY) layer security  utilizes the physical properties of wireless communication channels, such as  interference and channel fading, for ensuring perfectly secure communication  \cite{JR:Artifical_Noise1}--\nocite{JR:AN_MISO_secrecy,JR:rui_zhang}\cite{JR:Kwan_secure_imperfect}.  In particular, by exploiting  the extra degrees of freedom offered by multiple transmit antennas,
a properly designed artificial noise is transmitted concurrently with the information carrying signals to weaken the reception of the  eavesdroppers and to provide communication security. The authors of \cite{JR:Artifical_Noise1} and \cite{JR:AN_MISO_secrecy} investigated  secrecy capacity maximization  via power allocation and artificial noise transmission. The results in  \cite{JR:Artifical_Noise1} and \cite{JR:AN_MISO_secrecy}  indicate that a large amount of power is allocated  to artificial noise for providing secure communication which leads to a potent energy source in the RF. The notion  of  secure communication in energy harvesting systems has recently been pursued in  \cite{JR:rui_zhang} and \cite{JR:Kwan_secure_imperfect}. However, the resource allocation algorithms in \cite{JR:rui_zhang} and \cite{JR:Kwan_secure_imperfect}  were limited to the  case of a single information receiver and multiple single antenna eavesdroppers. In fact, optimal resource allocation for  secure communication  in WIPT systems with multiple desired information receivers and multiple multi-antenna eavesdroppers remains an unsolved and challenging problem.

In this paper, we address the above issues. To this end, we formulate the resource allocation algorithm design for secure multiuser  communication   with simultaneous WIPT as an optimization problem. The proposed problem formulation enables the dual use of artificial noise for facilitating  efficient wireless power transfer and guaranteeing communication security.  The resulting non-convex problem is recast as a semidefinite programming (SDP) which is solved optimally by SDP relaxation.


\section{System Model}
\label{sect:system model}

\subsection{Notation}
We use boldface capital and lower case letters to denote matrices and vectors, respectively. $\mathbf{A}^H$, $\Tr(\mathbf{A})$, $\Rank(\mathbf{A})$, and $\det(\mathbf{A})$ represent the Hermitian transpose, trace, rank, and determinant of  matrix $\mathbf{A}$; $\lambda_{\max}(\mathbf{A})$ denotes the maximum eigenvalue of matrix $\mathbf{A}$; $\mathbf{A}\succ \mathbf{0}$ and $\mathbf{A}\succeq \mathbf{0}$ indicate that $\mathbf{A}$ is a positive definite and a  positive semidefinite matrix, respectively; $\mathbf{I}_N$ is the $N\times N$ identity matrix; $\mathbb{C}^{N\times M}$ denotes the set of all $N\times M$ matrices with complex entries; $\mathbb{H}^N$ denotes the set of all $N\times N$ Hermitian matrices; the orthonormal null space of $\mathbf{A}\in\mathbb{C}^{M\times N}$ is defined as $\nullspace(\mathbf{A})
\triangleq\{\mathbf{y}\in\mathbb{C}^{N\times 1}:\mathbf{A}\mathbf{y} = \mathbf{0},\norm{\mathbf{y}}=1\}$.  A circularly symmetric complex Gaussian (CSCG) distribution is denoted by ${\cal CN}(\mathbf{m},\mathbf{\Sigma})$ with mean vector $\mathbf{m}$ and covariance matrix $\mathbf{\Sigma}$; $\sim$ indicates ``distributed as"; ${\cal E}\{\cdot\}$ denotes  statistical expectation; $\abs{\cdot}$ represents the absolute value of a complex scalar; $[x]^+$ stands for $\max\{0,x\}$.

\subsection{Channel Model}
\begin{figure}
 \centering
\includegraphics[width=3.5in]{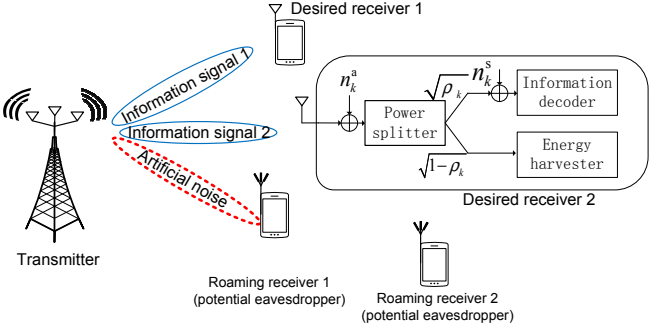}
 \caption{Multiuser downlink communication system model with $K=2$ single antenna desired receivers and $M=2$ roaming receivers. Each roaming receiver is equipped with $N_{\mathrm{R}}=2$ antennas.  }
 \label{fig:system_model}\vspace*{-2mm}
\end{figure}

We consider a multiuser downlink communication system with simultaneous WIPT. The system consists of one transmitter and two types of receivers, namely desired receivers and roaming receivers, cf. Figure \ref{fig:system_model}. The transmitter is equipped with $N_\mathrm{T}>1$ antennas serving  $K$ desired receivers and $M$ roaming receivers. The desired receivers are low computational capability  single antenna devices  which exploit the received signal powers in the RF for both information decoding and energy harvesting. On the other hand, each roaming receiver is equipped with $N_{\mathrm{R}}\ge 1$ antennas. We assume that $N_\mathrm{T}>N_\mathrm{R}$ and the roaming receivers are  wireless terminals from other communication systems searching for additional power supply in the RF. In particular, they temporally connect to the transmitter with the intend to harvest energy from the received signals radiated from the transmitter\footnote{A possible scenario of the considered system model is a cognitive radio setup. Specifically, the roaming receivers may be primary receivers which harvest energy from a secondary transmitter for extending the lifetime of the primary network.}. However, it is possible that the roaming receivers eavesdrop the information carrying signals deliberately. As a result, the $M$ roaming receivers are potential eavesdroppers which should be taken into account in the resource allocation algorithm design for providing secure communication.
We focus on a frequency flat fading channel and a time division duplexing (TDD) system. The transmitter can obtain perfect channel state information (CSI) of all receivers by exploiting channel reciprocity and handshaking signals. The received signals at the desired receivers and the roaming receiver are given by
\begin{eqnarray}
y_{k}&=&\mathbf{h}_k^H\mathbf{x}+n_k^{\mathrm{a}},\,\,  \forall k\in\{1,\dots,K\},\\
\mathbf{y}_{\mathrm{I}_m}&=&\mathbf{G}_m^H\mathbf{x}+\mathbf{n}_{\mathrm{a}_m},\,\,  \forall m\in\{1,\dots,M\},
\end{eqnarray}
where $\mathbf{x}\in\mathbb{C}^{N_{\mathrm{T}}\times1}$ denotes the transmitted signal vector. The channel vector between the transmitter and  desired receiver $k$ is denoted by $\mathbf{h}_k\in\mathbb{C}^{N_{\mathrm{T}}\times1}$. The channel  matrix between the transmitter and  roaming receiver $m$ is denoted by $\mathbf{G}_m\in\mathbb{C}^{N_{\mathrm{T}}\times N_{\mathrm{R}}}$. The channel vectors and matrices capture the joint effects of multipath fading and path loss. $n_k^{\mathrm{a}}\sim{\cal CN}(0,\sigma_{\mathrm{ant}}^2)$ and $\mathbf{n}_{\mathrm{a}_m}\sim{\cal CN}(\mathbf{0},\sigma_{\mathrm{ant}}^2\mathbf{I}_{N_{\mathrm{R}}})$ are additive white Gaussian noises (AWGN) caused by the thermal noises in the antennas of the desired receivers and the roaming receivers, respectively.

We assume a power splitting structure \cite{JR:WIP_receiver} is adopted in both the desired receivers and the roaming receivers. Specifically, desired receiver $k$ can split the received energy in the receiver RF front-end into
two power streams where $100\times\rho_k \%$ are used for decoding information and the remaining $100\times(1-\rho_k)\%$ are used for harvesting energy, cf. Figure \ref{fig:system_model}. Here, $0\leq\rho_k\leq1$ is the splitting ratio of desired receiver $k$. Similarly, power splitting is also performed at the roaming receivers for energy harvesting and information decoding. We assume that all receivers have enough energy for information decoding at the current time instant independent of the amount of harvested energy. The harvested energy is stored in a battery and used to support the normal operation of the receivers in the future.  Since a portion of received power is dedicated to energy harvesting, the equivalent receiving signal model for information decoding at desired receiver $k$ can be expressed as
\begin{eqnarray}
y_{k}^{\mathrm{ID}}&=&\sqrt{\rho_k}(\mathbf{h}_k^H\mathbf{x}+n_k^{\mathrm{a}})+n^{\mathrm{s}}_k,\,\,
\end{eqnarray}
where  $n^{\mathrm{s}}_k$ is AWGN with zero mean and variance $\sigma_{\mathrm{s}}^2$ caused by signal processing, cf. Figure \ref{fig:system_model}. We assume that the signal processing noise variances are the same for all receivers in this paper.

\subsection{Signal Model}
In each scheduling time slot, $K$ independent signal streams are transmitted simultaneously to $K$ desired receivers after  linear precoding. Specifically,  a dedicated beamforming vector, $\mathbf{w}_k\in\mathbb{C}^{N_{\mathrm{T}}\times1}$, is allocated to each desired receiver to facilitate information transmission.  On the other hand,  the  messages intended for the desired receiver  may be overheard by the roaming receivers since all receivers are in the range of service coverage. In order to guarantee communication security, artificial noise is transmitted currently with the information signals  for interfering the reception of the roaming receivers. As a result, the transmitted signal vector, $\mathbf{x}\in\mathbb{C}^{N_{\mathrm{T}}\times 1}$, is composed of the $K$ desired information signals and artificial noise, and can be expressed as
\begin{eqnarray}
\mathbf{x}&=&\sum_{k=1}^K\mathbf{w}_k s_k+\mathbf{v},
\end{eqnarray}
where $s_k\in\mathbb{C}$ is the signal intended for desired receiver $k$. Without loss of generality, we assume ${\cal E}\{\abs{s_k}^2\}=1,\forall k\in\{1,\ldots,K\}$. Variable $\mathbf{v}\in\mathbb{C}^{N_{\mathrm{T}}\times1}$ is the artificial noise vector  generated by the transmitter to degrade the quality of the signal received by the potential eavesdroppers. In particular, we model the artificial noise vector as $\mathbf{v}\sim\cal{CN}(\mathbf{0},\mathbf{V})$ with zero mean and covariance matrix $\mathbf{V}=\mathbf{v}\mathbf{v}^H$, $\mathbf{V}\in\mathbb{H}^{N_{\mathrm{T}}}, \mathbf{V}\succeq \mathbf{0}$.

\section{Resource Allocation Algorithm Design}
\label{sect:forumlation}
In this section, we first introduce the adopted quality of service (QoS)
metrics for the design of systems enabling efficient power transfer and secure communication. Then, the resource allocation algorithm design is formulated as a non-convex optimization problem and solved by SDP relaxation.

\subsection{Channel Capacity and Secrecy Capacity}
\label{subsect:Instaneous_Mutual_information}
The  channel capacity (bit/s/Hz) between the transmitter and desired receiver $k$ is given by
\begin{eqnarray}
C_{k}\hspace*{-1mm}&=&\hspace*{-1mm}\log_2(1+\Gamma_{k}),\quad \mbox{where}\\
\Gamma_{k}\hspace*{-1mm}&=&\hspace*{-1mm}\frac{\rho_k\abs{\mathbf{h}_k^H\mathbf{w}_k}^2}{\rho_k\Big(\sum\limits_
{\substack{j\neq k}}^K\abs{\mathbf{h}_k^H\mathbf{w}_j}^2+\Tr(\mathbf{h}_k\mathbf{h}_k^H\mathbf{V})+\sigma_{\mathrm{ant}}^2\Big)+\sigma_{\mathrm{s}}^2}
\end{eqnarray}
is the receive signal-to-interference-plus-noise ratio (SINR) at desired receiver $k$.

On the other hand, for guaranteeing  communication security,  the roaming receivers are treated as potential eavesdroppers who attempt to decode the messages transmitted for all $K$ desired receivers. Thereby, we focus on the worst case scenario. In particular, we assume that roaming receiver $m$ performs successive interference cancellation (SIC) to remove  all multiuser interference before decoding the message of receiver $k$. Therefore, the channel capacity between the transmitter and roaming receiver $m$ for decoding the signal of desired receiver $k$ can be represented as
\begin{eqnarray}\label{eqn:Capacity_eve}
\hspace*{-6mm}C_{\mathrm{I}_{m,k}}\hspace*{-2mm}&=&\hspace*{-2mm}\log_2\det(\mathbf{I}_{N_{\mathrm{R}}}\hspace*{-0.5mm}+\hspace*{-0.5mm}
\mathbf{\Delta}_{m}^{-1}
\rho_{\mathrm{E}_m}\mathbf{G}_m^H\mathbf{w}_k\mathbf{w}_k^H\mathbf{G}_m),\\
\mathbf{\Delta}_{m}\hspace*{-2mm}&=&\hspace*{-2mm}\rho_{\mathrm{E}_m}\mathbf{\Sigma}_{m}+
\sigma_\mathrm{s}^2\mathbf{I}_{N_{\mathrm{R}}},\quad\mbox{and}\\ \label{eqn:Sigma}
\mathbf{\Sigma}_{m}\hspace*{-2mm}&=&\hspace*{-2mm}\mathbf{G}_m^H\mathbf{V}\mathbf{G}_m+\sigma_{\mathrm{ant}}^2\mathbf{I}_{N_{\mathrm{R}}},
\end{eqnarray}
where $0\le \rho_{\mathrm{E}_m}\le 1$ is the power splitting ratio and $\mathbf{\Sigma}_{m}$ is the interference-plus-noise covariance matrices for roaming receiver (potential eavesdropper) $m$. In practice, the roaming receiver can be malicious and devote all the received energy to information decoding. Thus, the channel capacity in (\ref{eqn:Capacity_eve}) is bounded above by
\begin{eqnarray}\label{eqn:Capacity_eve_up}
C_{\mathrm{I}_{m,k}}^{\mathrm{up}}=\log_2\det(\mathbf{I}_{N_{\mathrm{R}}}\hspace*{-0.5mm}+\hspace*{-0.5mm}(\mathbf{\Sigma}_{m}+\sigma_\mathrm{s}^2\mathbf{I}_{N_{\mathrm{R}}})^{-1}
\mathbf{G}_m^H\mathbf{w}_k\mathbf{w}_k^H\mathbf{G}_m)
\end{eqnarray}
which is obtained by setting $\rho_{\mathrm{E}_m}=1$ in  (\ref{eqn:Capacity_eve}).

 Consequently, the maximum achievable secrecy capacity of desired receiver $k$ under the considered worst case scenario is given by
\begin{eqnarray}\label{eqn:secure_cap}
C_{\mathrm{sec}_k}&=&\Big[C_{k}-\underset{\forall m}{\max}\,\{ C_{\mathrm{I}_{m,k}}^{\mathrm{up}}\}\Big]^+.
\end{eqnarray}
\begin{Remark}
 We note that the results of this work are also applicable  to the case of roaming receivers (potential eavesdroppers) employing single user detectors by modifying the term $\mathbf{\Sigma}_{m}$ in (\ref{eqn:Sigma}) accordingly.
\end{Remark}

\subsection{Energy Harvesting}
\label{subsect:harvested energy}
For transferring power\footnote{ In this paper, we study the algorithm design for a normalized unit energy, i.e., Joule-per-second. Thus, the terms ``energy" and ``power" are interchangeable under this context.} to both desired receivers and roaming receivers, both the information signal, $\mathbf{w}_ks_k,\forall k\in\{1,\ldots,K\}$, and the artificial noise, $\mathbf{v}$, play an important role in the system design. In particular, they act as energy harvesting sources for the receivers. The total amount of energy harvested by desired receiver $k$ is given by
\begin{eqnarray}
E_{k}=\eta(1-\rho_k)\Big(\sum_{j=1}^K\abs{\mathbf{h}_k^H\mathbf{w}_j}^2+\Tr(\mathbf{h}_k\mathbf{h}_k^H\mathbf{V})+\sigma_{\mathrm{ant}}^2\Big),
\end{eqnarray}
where $0\leq\eta\leq1$ denotes the efficiency for converting the received RF energy to electrical energy for storage. We assume that it is a constant and is identical for all receivers.

Similarly, the total amount of energy harvested by roaming receiver $m$ is given by
\begin{eqnarray}
E_{\mathrm{I}_m}&=&\eta_m(1-\rho_{\mathrm{E}_m})\Big(\sum_{k=1}^K\Tr(\mathbf{G}_m^H\mathbf{w}_k\mathbf{w}_k^H\mathbf{G}_m)\nonumber\\
&+&\Tr(\mathbf{G}_m\mathbf{G}_m^H\mathbf{V})+N_{\mathrm{R}}\sigma_{\mathrm{ant}}^2\Big).
\end{eqnarray}

\subsection{Optimization Problem Formulation}
\label{sect:cross-Layer_formulation}
The system objective is to minimize the total transmit power while providing QoS with regard to communication security and power transfer. The resource allocation algorithm design  is formulated as an optimization problem which is given by
\begin{eqnarray} \label{eqn:cross-layer}
&&\hspace*{-2mm} \underset{ \mathbf{V}\in \mathbb{H}^{N_{\mathrm{T}}},\mathbf{w}_k,\rho_k}{\mino}\,\, \sum_{k=1}^K\norm{\mathbf{w}_k}^2+\Tr(\mathbf{V})\nonumber\\
\notag \mbox{s.t.} &&\hspace*{-5mm}\mbox{C1: }\Gamma_{k}\ge \Gamma_{\mathrm{req},k},\,\, \forall k, \notag\\
&&\hspace*{-5mm}\mbox{C2: }C_{\mathrm{I}_m}^{\mathrm{up}}\le R_{\mathrm{eav}_{m,k}},\,\, \forall k,\forall m,\notag\\
&&\hspace*{-5mm}\mbox{C3: }E_{k}\ge P^{\mathrm{min}}_{k},\,\, \forall k,\notag\\
&&\hspace*{-5mm}\mbox{C4:
}\notag\eta\Big(\sum_{k=1}^K\Tr(\mathbf{G}_m^H\mathbf{w}_k\mathbf{w}_k^H\mathbf{G}_m)+\Tr(\mathbf{G}_m\mathbf{G}_m^H\mathbf{V})\\
&&\hspace*{2mm}+N_{\mathrm{R}}\sigma_{\mathrm{ant}}^2\Big)\ge P^{\min}_{\mathrm{I}_m},\,\, \forall m, \notag\\
&&\hspace*{-5mm}\mbox{C5:}\,\,  0\leq\rho_k\leq1,\,\, \forall k, \quad\mbox{C6:}\,\, \mathbf{V}\succeq \mathbf{0}.
\end{eqnarray}
Constraint C1 indicates that the receive SINR at desired receiver $k$ is required to be larger than a given threshold, $\Gamma_{\mathrm{req},k}>0$.  Since any desired receiver could  be chosen as an eavesdropping target of roaming receiver $m$,
the upper limit $R_{\mathrm{eav}_{m,k}}$ is imposed in C2 to restrict the channel capacity of roaming receiver $m$ if it  attempts to decode the message of desired receiver $k,\forall k$.  Notice that in practice we are interested in the case of  $C_{k}> R_{\mathrm{eav}_{m,k}},\forall k,\forall m$, for ensuring secure communication, i.e., $C_{\mathrm{sec}_k} \ge C_{k}-\underset{\forall m}{\max}\,\{R_{\mathrm{eav}_{m,k}}\}= \log_2(1+\Gamma_{\mathrm{req},k})-\underset{\forall m}{\max}\,\{R_{\mathrm{eav}_{m,k}}\}>0$. In particular, the parameters
$\Gamma_{\mathrm{req},k}$ and $R_{\mathrm{eav}_{m,k}}$ can be selected to provide flexibility in designing power efficient resource allocation algorithms for different applications.  Constants $P_k^{\min}$ and $P_{\mathrm{I}_m}^{\min}$ in constraints C3 and C4 specify the minimum required energy harvested at  desired receiver $k$ and roaming receiver $m$, respectively. The physical meaning of constraint C4 is that the transmitter only guarantees the minimum required  harvested power at roaming receiver $m$ if it does not attempt to eavesdrop, i.e., $\rho_{\mathrm{E}_m}=0$. Constraint C5 specifies the physical constraints of the power splitter. In particular, we assume that the power splitter is a passive device which does not consume any received signal power in splitting the received signal power. Besides, no extra power can be gained by splitting power.  Constraint C6 and $\mathbf{V}\in \mathbb{H}^{N_{\mathrm{T}}}$ ensure that the covariance matrix $\mathbf{V}$ is  a positive semidefinite Hermitian matrix.

\subsection{Optimization Solution}\label{sect:solution}
It can be observed that optimization problem (\ref{eqn:cross-layer}) is non-convex due to constraints C1 and C2. To overcome the non-convexity of C2,  we recast the considered problem using SDP. We first replace $\mathbf{w}_k\mathbf{w}_k^H$ in (\ref{eqn:cross-layer}) with $\mathbf{W}_k=\mathbf{w}_k\mathbf{w}_k^H$ and rewrite C2 as
\begin{eqnarray}\label{eqn:original_c2}
\mbox{C2:}\,\,\hspace*{-5mm} &&\det\hspace*{-0.5mm}\Big[\mathbf{I}_{N_{\mathrm{R}}}\hspace*{-0.5mm}+ \hspace*{-0.5mm} \mathbf{Q}_{m}^{-1}\mathbf{G}_m^H\mathbf{W}_k\mathbf{G}_m\Big]\hspace*{-0.5mm}\le\hspace*{-0.5mm}\xi_{\mathrm{eav}_{m,k}},\, \forall m,k, \,\, \\
\notag\mathbf{Q}_{m}\,\,\hspace*{-5mm}&&=\mathbf{G}_m^H\mathbf{V}\mathbf{G}_m+
(\sigma_{\mathrm{ant}}^2\hspace*{-0.5mm}+\hspace*{-0.5mm}\sigma_{\mathrm{s}}^2)\mathbf{I}_{N_{\mathrm{R}}}\hspace*{-0.5mm}\succ\hspace*{-0.5mm} \mathbf{0},
\end{eqnarray} where $\xi_{\mathrm{eav}_{m,k}}=2^{R_{\mathrm{eav}_{m,k}}}$,  $\xi_{\mathrm{eav}_{m,k}}> 1$
for $R_{\mathrm{eav}_{m,k}}> 0$, is an auxiliary constant.  Then, we introduce the following proposition for simplifying the considered optimization problem.
\begin{proposition}\label{prop:relaxed_c2} For $R_{\mathrm{eav}_{m,k}}> 0,\forall m,k$, the following implication on constraint C2 holds:
\begin{eqnarray}\label{eqn:det_to_matrix}\notag
\mbox{C2}\Rightarrow\overline{\mbox{C2}}\mbox{: } \mathbf{G}_m^H\mathbf{W}_k\mathbf{G}_m\preceq (\xi_{\mathrm{eav}_{m,k}} -1)\mathbf{Q}_{m},\,\, \forall m,k,
\end{eqnarray}
\end{proposition}
where $\overline{\mbox{C2}}$ is a linear matrix inequality (LMI) constraint. In particular, constraints $\overline{\mbox{C2}}$ and ${\mbox{C2}}$ are equivalent if $\Rank(\mathbf{W}_k)=1,\forall k$.

\,\,\emph{Proof:} Please refer to Appendix A for the proof of Proposition \ref{prop:relaxed_c2}.

Now, we apply Proposition \ref{prop:relaxed_c2} to (\ref{eqn:cross-layer}) by replacing constraint $\mbox{C2}$ with constraint $\overline{\mbox{C2}}$.  Then, the new optimization problem under the SDP reformulation can be written as
\begin{eqnarray}\label{eqn:rank_one}
&&\hspace*{2mm} \underset{\mathbf{W}_k, \mathbf{V}\in \mathbb{H}^{N_{\mathrm{T}}}, \rho_k}{\mino}\,\, \sum_{k=1}^K\Tr(\mathbf{W}_k)+\Tr(\mathbf{V})\notag\\
\mbox{s.t.} &&\hspace*{-5mm}\mbox{C1: }\frac{1}{\Gamma_{\mathrm{req},k}}\Tr(\mathbf{h}_k\mathbf{h}_k^H\mathbf{W}_k)-\sum\limits_
{\substack{j\neq k}}^K\Tr(\mathbf{h}_k\mathbf{h}_k^H\mathbf{W}_j)\notag \\
&&-\Tr(\mathbf{h}_k\mathbf{h}_k^H\mathbf{V})\ge \sigma_{\mathrm{ant}}^2+\frac{1}{\rho_k}\sigma_{\mathrm{s}}^2,\,\, \forall k, \notag\\
&&\hspace*{-5mm}\overline{\mbox{C2}}\mbox{: } \notag\mathbf{G}_m^H\mathbf{W}_k\mathbf{G}_m\preceq (\xi_{\mathrm{eav}_{m,k}} -1)\mathbf{Q}_{m},\,\, \forall m,k,\notag\\
&&\hspace*{-5mm}\mbox{C3: }\notag\Tr(\mathbf{h}_k\mathbf{h}_k^H(\mathbf{V}+\sum_{j=1}^K\mathbf{W}_j))\ge \frac{P^{\mathrm{min}}_{k}}{\eta(1-\rho_k)}-\sigma_{\mathrm{ant}}^2,\forall k,\notag\\
&&\hspace*{-5mm}\mbox{C4:
}\notag\Tr(\mathbf{G}_m^H(\mathbf{V}+\sum_{k=1}^K\mathbf{W}_k)\mathbf{G}_m)\ge \frac{P^{\min}_{\mathrm{I}_m}}{\eta}-N_{\mathrm{R}}\sigma_{\mathrm{ant}}^2,\,\, \forall m,\notag\\
&&\hspace*{-5mm}\mbox{C5:}\,\,  0\leq\rho_k\leq1,\,\, \forall k, \,\,\,
\mbox{C6:}\,\, \mathbf{V}\succeq \mathbf{0},\notag\\
&&\hspace*{-5mm}\mbox{C7:}\,\, \mathbf{W}_k\succeq \mathbf{0},\,\, \forall k, \quad\,\,\,
\mbox{C8:}\,\, \Rank(\mathbf{W}_k)=1,\,\, \forall k.
\end{eqnarray}
Constraints C7, C8, and $\mathbf{W}_k\in\mathbb{H}^{N_{\mathrm{T}}},\forall k$, are imposed to guarantee that $\mathbf{W}_k=\mathbf{w}_k\mathbf{w}_k^H$ holds after optimization. In general, replacing constraint
C2  by $\overline{\mbox{C2}}$  leads to a larger feasible solution set for optimization, cf. Proposition \ref{prop:relaxed_c2}.  However, the optimization problems in (\ref{eqn:cross-layer}) and (\ref{eqn:rank_one}) are equivalent  for $\Rank(\mathbf{W}_k)= 1$, $\forall k$. Thus, in the sequel, we focus on the new optimization problem in (\ref{eqn:rank_one}).

 Although the new constraint $\overline{\mbox{C2}}$ is an affine function with respect to the optimization variables,  it can be verified that the problem in (\ref{eqn:rank_one}) is still non-convex due to the combinatorial rank constraint in C8. For facilitating an efficient design of the resource allocation algorithm, we adopt a SDP relaxation approach. Specifically, we relax constraint $\mbox{C8: }\Rank(\mathbf{W})=1$, i.e., we remove it from the problem formulation, such that the considered problem becomes a convex SDP.  The SDP relaxed problem formulation of (\ref{eqn:rank_one}) is given by
\begin{eqnarray}
\label{eqn:sdp_relaxation}&&\hspace*{-20mm}\underset{\mathbf{W}_k, \mathbf{V}\in \mathbb{H}^{N_{\mathrm{T}}}, \rho_k}{\mino}\,\, \sum_{k=1}^K\Tr(\mathbf{W}_k)+\Tr(\mathbf{V})\notag\\
\mbox{s.t.} &&\hspace*{-5mm}\mbox{C1},\,\overline{\mbox{C2}},\,\mbox{C3},\,\mbox{C4},\,\mbox{C5},\,\mbox{C6},\,\mbox{C7}.
\end{eqnarray}
We note that the relaxed problem in (\ref{eqn:sdp_relaxation}) can be solved efficiently by numerical solvers such as CVX \cite{website:CVX}. If the obtained solution $\mathbf{W}_k$ for (\ref{eqn:sdp_relaxation}) admits a rank-one matrix, then the problems in (\ref{eqn:cross-layer}), (\ref{eqn:rank_one}), and (\ref{eqn:sdp_relaxation}) share the same optimal solution and the same optimal objective value.

Now, we introduce the following  theorem for revealing the tightness of the SDP relaxation adopted in  (\ref{eqn:sdp_relaxation}).
\begin{Thm}\label{thm:rankone_condition} Suppose the optimal solution (\ref{eqn:sdp_relaxation}) is denoted by $\{\mathbf{W}_k^*,\mathbf{V}^*,\rho_k^*\}$, ${\Gamma}_{\mathrm{req},k}>0$, and $R_{\mathrm{eav}_{m,k}}>0$. If  $\exists k:\Rank(\mathbf{W}^*_k)>1$, then we can construct another solution of (\ref{eqn:sdp_relaxation}), denoted as  $\{\mathbf{\widetilde  W}_k ,\mathbf{\widetilde  V},\widetilde \rho_k\}$, which not only achieves the same objective value as $\{\mathbf{W}_k^*,\mathbf{V}^*,\rho_k^*\}$, but also admits a rank-one matrix, i.e.,  $\Rank(\mathbf{\widetilde W}_k)=1,\forall k$.
\end{Thm}
\emph{\quad Proof: } Please refer to Appendix B for a proof of Theorem \ref{thm:rankone_condition} and a method for constructing $\{\mathbf{\widetilde  W}_k ,\mathbf{\widetilde  V},\widetilde \rho_k\}$ with $\Rank(\mathbf{\widetilde W}_k)=1,\forall k$.

In other words, by applying Theorem \ref{thm:rankone_condition} and Proposition \ref{prop:relaxed_c2}, the global optimal solution of (\ref{eqn:cross-layer}) is obtained.

\section{Results}
In this section, we study the system performance of the optimal resource allocation design via simulation. In particular, we solve the optimization problem in (\ref{eqn:cross-layer}) for  different channel realizations and show the corresponding average system performance.  We adopt the TGn path loss model \cite{report:tgn} with transmit and receive antenna gains of 10 dB. In particular, we  assume a carrier center frequency of $470$ MHz \cite{report:80211af}. There are $K=3$ desired receivers and $M=2$ roaming receivers (potential eavesdroppers),  which are uniformly distributed in the range between a reference distance of 2 meters and a maximum distance of 50 meters. Each roaming receiver is equipped with $N_{\mathrm{R}}=2$ antennas. The multipath fading coefficients are modeled as independent and identically distributed Rician fading with Rician factor 3 dB. We set the minimum required SINRs of all desired receivers to $\Gamma_{\mathrm{req},k}=\Gamma_{\mathrm{req}},\forall k\in\{1,\ldots,K\}$, the maximum data rate tolerance of each roaming receiver is $R_{\mathrm{eav}_{m,k}}=1$ bit/s/Hz, $\forall m,k$, and the minimum required harvested power  for all receivers is $P^{\mathrm{min}}_{k}=P^{\min}_{\mathrm{I}_m}=0$ dBm.  The energy conversion efficiency in converting RF energy to electrical energy is  $\eta=0.5$. The antenna noise power is $\sigma_{\mathrm{ant}}^2= -114$ dBm at a temperature of 290 Kelvin. We assume that a 8-bit uniform quantizer is employed in the analog-to-digital converter at the analog front-end of each receiver leading to a signal processing noise of $\sigma_{\mathrm{s}}^2= -53$ dBm.
\begin{figure}[t]
 \centering
\includegraphics[width=3.5in]{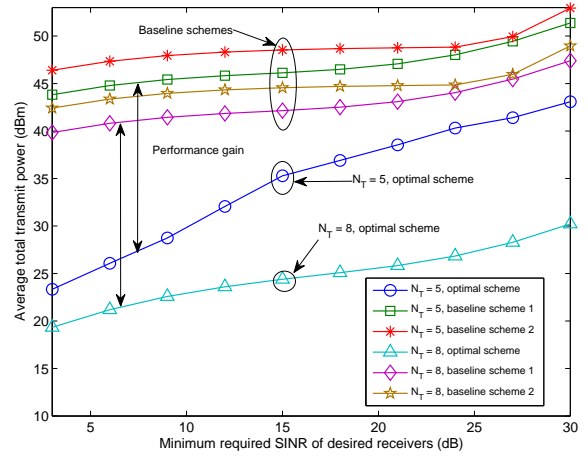}\vspace*{-1mm}
 \caption{Average total transmit power (dBm) versus the minimum required SINR of the desired receivers, $\Gamma_{\mathrm{req}}$,  for $N_{\mathrm{T}}=5$ and $N_{\mathrm{T}}=8$. The double-sided arrows indicate the power gains achieved by the optimal scheme compared to the baseline schemes.}\vspace*{-4mm}
 \label{fig:pt_SINR}
\end{figure}
\subsection{Average Total Transmit Power}
\begin{figure}[t]
 \centering
\includegraphics[width=3.5in]{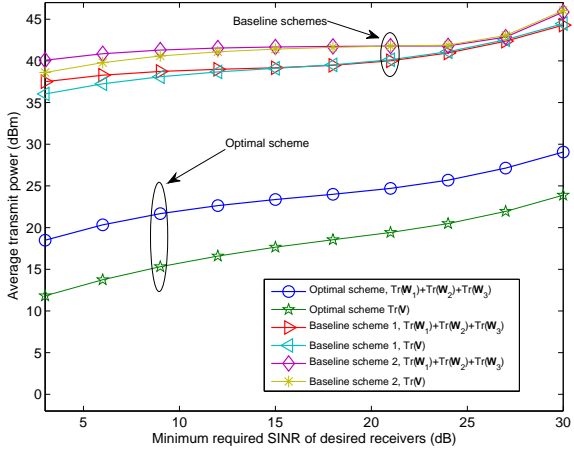}\vspace*{-1mm}
 \caption{Average transmit power allocation (dBm) of desired signals and artificial noise versus the minimum required SINR of the desired receivers, $\Gamma_{\mathrm{req}}$,  for $N_{\mathrm{T}}=8$.}
 \label{fig:sig_AN}\vspace*{-5mm}
\end{figure}

In Figure \ref{fig:pt_SINR}, we study the average total transmit power of the optimal scheme versus  the minimum required SINR,  $\Gamma_{\mathrm{req}}$, for different numbers of transmit antennas  and different resource allocation schemes. It can be observed that the total transmit power increases monotonically with an increasing value of $\Gamma_{\mathrm{req}}$.  The reason behind this is twofold. First, a higher transmit power for information signals, $\mathbf{w}_ks_k,\forall k$, is required to satisfy the increasingly stringent requirement on $\Gamma_{\mathrm{req},k}$. Second, a higher amount of power also has to be allocated to the artificial noise, $\mathbf{v}$, for neutralizing the increased information leakage due to the higher  power of  $\mathbf{w}_ks_k,\forall k$, cf.
Figure \ref{fig:sig_AN}. On the other hand,  it can be observed that a significant power saving can be achieved by the proposed optimal scheme  when the number of antennas increase from $N_{\mathrm{T}}=5$ to $N_{\mathrm{T}}=8$. This is due to the fact that the degrees of freedom for resource allocation increase if the number of transmit antennas increases, which enables a more power efficient resource allocation.

For comparison, we also show  the performance of two simple suboptimal baseline schemes. For baseline scheme 1, zero-forcing beamforming is performed for the desired signals  such that the desired receivers do not experience any multiuser interference. In particular, we calculate the eigenvalue decomposition of $\mathbf{H}_{-k}\mathbf{H}_{-k}^H=\mathbf{U}_k\mathbf{\Sigma}_k\mathbf{U}^H_k$ for desired receiver $k$ where $\mathbf{H}_{-k}=[\mathbf{h}_1\,\ldots\,\mathbf{h}_{k-1}\,\mathbf{h}_{k+1}\,\ldots\,\mathbf{h}_K]$, $\mathbf{U}_k$ and $\mathbf{\Sigma}_k$  are an $N_\mathrm{T}\times N_\mathrm{T}$ unitary matrix and a diagonal matrix with ascending eigenvalues of  $\mathbf{H}_{-k}\mathbf{H}_{-k}^H$ as main diagonal elements, respectively. Then, we select $\mathbf{W}_k=q_{\mathrm{sub}_k}\mathbf{w}_{\mathrm{sub}_k}\mathbf{w}_{\mathrm{sub}_k}^H$, where $q_{\mathrm{sub}_k}\ge 0$ is a new scalar optimization variable and  $\mathbf{w}_{\mathrm{sub}_k}$ is the first column vector\footnote{In general, different column vectors with respect to the null space of $\mathbf{H}_{-k}\mathbf{H}_{-k}^H$ can be used as zero-forcing beamforming vector.  For algorithm computational simplicity, we select the first column vector corresponding to the minimum eigenvalue of matrix $\mathbf{H}_{-k}\mathbf{H}_{-k}^H$ as zero-forcing beamforming vector. } of $\mathbf{U}_k$ such that $\mathbf{H}_{-k}^H\mathbf{w}_{\mathrm{sub}_k}=\mathbf{0}$. In other words, the directions of the beamforming matrices are fixed for all desired users. Then, we minimize the total transmit power by optimizing  $q_{\mathrm{sub}_k},\mathbf{V}$, and $\rho_k$ subject to the same constraints as in (\ref{eqn:sdp_relaxation}). We note that the zero-forcing beamforming matrix admits a rank-one structure.    As for baseline scheme 2, it shares the same resource allocation policy  as baseline scheme 1 except that we set $\rho_k=0.5, \forall k$. It can be observed in Figure \ref{fig:pt_SINR} that the optimal scheme achieves  significant  power savings over the two baseline schemes. Notably, the performance gain of  the optimal scheme over the two baseline schemes  is further enlarged for an increasing number of transmit antennas  $N_{\mathrm{T}}$. This can be explained by the fact that the optimal scheme can fully utilize the degrees of freedom offered by the system for resource allocation. In contrast,  although multiuser interference is eliminated  in the two baseline schemes, the degrees of freedom for resource allocation in the baseline schemes are limited which results in a higher transmit power. Furthermore, the performance gap between baseline scheme 1 and baseline scheme 2 reveals the performance gain in the baseline schemes due to the optimization of the power splitting ratio $\rho_k,\forall k$.

Figure \ref{fig:sig_AN} depicts the average transmit power allocated to the desired information signals and the artificial noise, i.e., $\Tr(\mathbf{W}_1)+\Tr(\mathbf{W}_2)+\Tr(\mathbf{W}_3)$ and $\Tr(\mathbf{V})$, for $N_{\mathrm{T}}=8$.  It
can be seen that the powers allocated to both the
information signal and the artificial noise increase rapidly
with increase of minimum SINR requirement $\Gamma_{\mathrm{req}}$. Besides,  both the optimal scheme and the two baseline schemes indicate that a large portion of the total transmit power  is allocated to the artificial noise. These results suggest that artificial noise generation is crucial for  guaranteeing communication security and providing efficient wireless power transfer.

\subsection{Secrecy Capacity}
\begin{figure}[t]
 \centering
\includegraphics[width=3.5in]{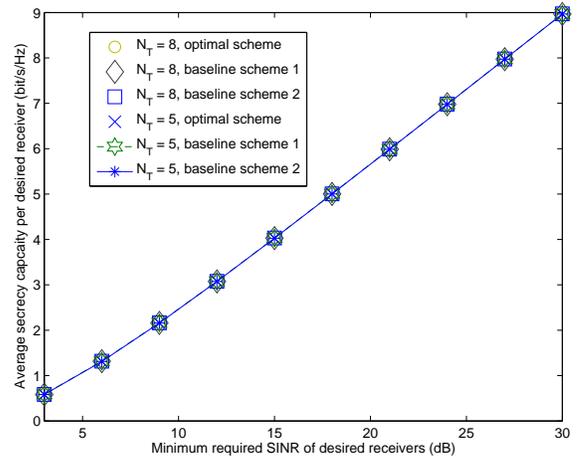}\vspace*{-1mm}
 \caption{Average secrecy capacity per desired receiver (bit/s/Hz) versus the minimum required SINR  of the desired receivers, $\Gamma_{\mathrm{req}}$.}
 \label{fig:cap_SINR}\vspace*{-3mm}
\end{figure}

Figure \ref{fig:cap_SINR} plots the average secrecy capacity per desired receiver  with respect to the minimum required SINR $\Gamma_{\mathrm{req}}$ of the desired receivers for different numbers of transmit antennas and different resource allocation schemes.  It can be seen that the average system secrecy capacity, i.e., $C_{\mathrm{sec}_k}$, increases with  $\Gamma_{\mathrm{req}}$
since the channel capacity of roaming receiver (potential eavesdroppers) $m$  is limited to $R_{\mathrm{eav}_{m,k}}=1$ bit/s/Hz. Besides, all  considered schemes are able to guarantee the QoS requirement on communication security (constraints C1 and C2) and achieve the same value of secrecy capacity. However, the two baseline schemes achieve the same secrecy capacity as the optimal scheme at the expense of a significantly higher transmit power, cf. Figure \ref{fig:pt_SINR}.

\begin{figure}[t]
 \centering
\includegraphics[width=3.5in]{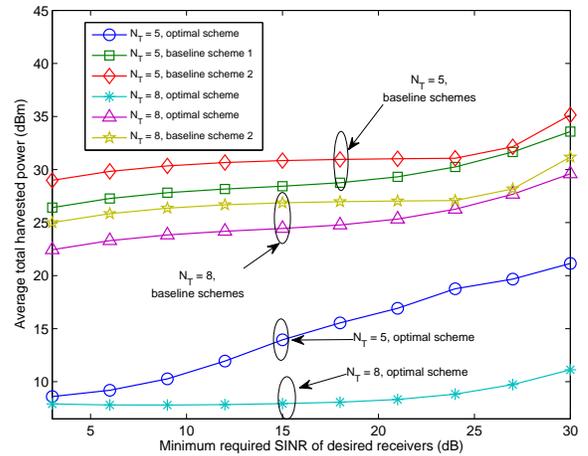}\vspace*{-1mm}
 \caption{Average total harvested power (dBm) versus the minimum required SINR  of the desired receivers, $\Gamma_{\mathrm{req}}$.}
 \label{fig:hp_SINR}\vspace*{-5mm}
\end{figure}

\subsection{Average Total Harvested Power}
In Figure \ref{fig:hp_SINR}, we study
the average total harvested power versus the minimum required SINR $\Gamma_{\mathrm{req}}$ of the desired receivers for different resource allocation schemes and different numbers of transmit antennas. The average total
harvested power  is computed by assuming the roaming receivers (potential eavesdroppers) do not eavesdrop.
 It is expected that the total
average harvested powers of all resource allocation schemes increase with $\Gamma_{\mathrm{req}}$ as more energy is available in the RF for an increasing $\Gamma_{\mathrm{req}}$, cf. Figure \ref{fig:pt_SINR}. Besides, the receivers for the two baseline schemes are able to harvest more power from the RF compared to the optimal scheme. The superior energy harvesting performance  of the baseline schemes compared to the optimal scheme comes at the expense of an exceedingly  large  transmit power. On the other hand, it can be observed that the average
total harvested power in the system decreases with an increasing number of transmit antennas.  Indeed, the extra degrees of freedom offered by the increasing number of antennas improve the efficiency of resource allocation. In particular,  the information leakage can be efficiently  reduced and artificial noise jamming can be more accurately performed. As a results, a  lower amount of transmit power is required to fulfill all QoS requirements and a lower amount of power is harvested from the RF.

\section{Conclusions}\label{sect:conclusion}
In this paper, we studied the power efficient resource allocation algorithm design for secure multiuser communication systems with simultaneous information and power transfer. The algorithm design was formulated as a non-convex optimization problem which took into account the QoS requirements on communication security and efficient power transfer. We applied  SDP relaxation to obtain the optimal solution. Simulation results confirmed the remarkable performance of our proposed optimal resource allocation scheme. Ensuring  secure communication and efficient power transfer for   multiple antennas desired receivers  is an interesting topic for future research.

\section*{Appendix}
\subsection{Proof of Proposition \ref{prop:relaxed_c2}}
We start the proof by  expressing  constraint C2 as
\begin{eqnarray}\label{eqn:det_ineq}
\det(\mathbf{I}_{N_{\mathrm{R}}}+\mathbf{Q}_{m}^{-1}\mathbf{G}_m^H\mathbf{W}_k\mathbf{G}_m)\hspace*{-3mm}&\le&\hspace*{-3mm}\xi_{\mathrm{eav}_{m,k}}\\
\stackrel{(a)}{\Longleftrightarrow }\det(\mathbf{I}_{N_{\mathrm{R}}}+\mathbf{Q}_{m}^{-1/2}\mathbf{G}_m^H\mathbf{W}_k\mathbf{G}_m\mathbf{Q}_{m}^{-1/2})
\hspace*{-3mm}&\le&\hspace*{-3mm}\xi_{\mathrm{eav}_{m,k}},\label{eqn:det_ineq2}
\end{eqnarray}
where $(a)$ is due to the fact that $\det(\mathbf{I}+\mathbf{AB})=\det(\mathbf{I}+\mathbf{BA})$ holds for   any matrices $\mathbf{A}$ and $\mathbf{B}$. Then, we introduce the following lemma which provides a lower bound on the left hand side of (\ref{eqn:det_ineq2}).
\begin{Lem}\label{lemma:det_trace} For any square matrix $\mathbf{A}\succeq \mathbf{0}$, we have the following inequality \cite{JR:AN_MISO_secrecy}:
\begin{eqnarray}
\det(\mathbf{I}+\mathbf{A})\ge 1+\Tr(\mathbf{A}),
\end{eqnarray}
\end{Lem}
where the equality holds if and only if $\Rank(\mathbf{A})\le 1$.

Exploiting Lemma \ref{lemma:det_trace}, the left hand side of (\ref{eqn:det_ineq2}) is bounded below by
\begin{eqnarray}&&\det(\mathbf{I}_{N_{\mathrm{R}}}+\mathbf{Q}_{m}^{-1/2}\mathbf{G}_m^H\mathbf{W}_k\mathbf{G}_m\mathbf{Q}_{m}^{-1/2}) \notag\\
&{\ge}& 1+\Tr(\mathbf{Q}_{m}^{-1/2}\mathbf{G}_m^H\mathbf{W}_k\mathbf{G}_m\mathbf{Q}_{m}^{-1/2}).\label{eqn:trace_ineq3}
\end{eqnarray}

Subsequently, by combining equations (\ref{eqn:det_ineq}), (\ref{eqn:det_ineq2}),  and (\ref{eqn:trace_ineq3}), we have the following implications:
\begin{subequations}
\begin{eqnarray}\notag
\hspace*{-5mm}&&\mbox{(\ref{eqn:det_ineq})} \Longleftrightarrow \mbox{(\ref{eqn:det_ineq2})}\notag\\
\hspace*{-5mm}&\Longrightarrow& \hspace*{-2mm}\Tr(\mathbf{Q}_{m}^{-1/2}\mathbf{G}_m^H\mathbf{W}_k\mathbf{G}_m\mathbf{Q}_{m}^{-1/2})\le \xi_{\mathrm{eav}_{m,k}} -1\\
\hspace*{-5mm}&\stackrel{(b)}{ \Longrightarrow }&\hspace*{-2mm} \lambda_{\max}(\mathbf{Q}_{m}^{-1/2}\mathbf{G}_m^H\mathbf{W}_k\mathbf{G}_m\mathbf{Q}_{m}^{-1/2})\le \xi_{\mathrm{eav}_{m,k}} -1\\
\hspace*{-5mm}&\Longleftrightarrow&\hspace*{-2mm}\mathbf{Q}_{m}^{-1/2}\mathbf{G}_m^H\mathbf{W}_k\mathbf{G}_m\mathbf{Q}_{m}^{-1/2} \preceq (\xi_{\mathrm{eav}_{m,k}} -1)\mathbf{I}_{N_{\mathrm{R}}}\\
\hspace*{-5mm}&\Longleftrightarrow &\hspace*{-2mm} \mathbf{G}_m^H\mathbf{W}_k\mathbf{G}_m\preceq (\xi_{\mathrm{eav}_{m,k}} -1)\mathbf{Q}_{m}, \label{eqn:trace_final}
\end{eqnarray}
\end{subequations}
where $(b)$ is due to  $\Tr(\mathbf{A})\ge \lambda_{\mathrm{max}}(\mathbf{A})$ for a positive semidefinite matrix $\mathbf{A}\succeq \mathbf{0}$. We note that  equations (\ref{eqn:det_ineq}) and (\ref{eqn:trace_final}) are equivalent when $\Rank(\mathbf{W}_k)= 1,\forall k$.

\subsection{Proof of Theorem \ref{thm:rankone_condition}}
We follow a similar approach as in \cite{JR:rui_zhang,JR:Kwan_secure_imperfect} to prove Theorem 1. The proof is divided into two parts. In the first part, we study the solution structure of (\ref{eqn:sdp_relaxation}). Then in the second part, we propose a simple method for constructing an optimal solution with rank-one $\mathbf{W}_k$. In order to verify the tightness of the adopted SDP relaxation,  we analyze the Karush-Kuhn-Tucker (KKT) conditions of the SDP relaxed optimization problem in (\ref{eqn:sdp_relaxation}) by first introducing the corresponding  Lagrangian and the dual problem. The
Lagrangian of (\ref{eqn:sdp_relaxation}) can be expressed as
\begin{eqnarray}
\hspace*{-2mm}&&{\cal L}\Big(\mathbf{W}_k,\mathbf{V},\rho_k,\mathbf{Z}_k,\mathbf{Y},\mathbf{X}_{m,k},\beta_k,\alpha_k,\nu_m\Big)\\
\hspace*{-5mm}&=&\hspace*{-3mm}\sum_{k=1}^K\Tr(\mathbf{W}_k)+\Tr(\mathbf{V})-\Tr(\mathbf{Y}\mathbf{V})-\sum_{k=1}^K\Tr(\mathbf{Z}_k\mathbf{W}_k)\notag\\
\hspace*{-5mm}&+&\hspace*{-3mm}\sum_{k=1}^K\alpha_k\Big[-\frac{1}{\Gamma_{\mathrm{req},k}}\Tr(\mathbf{h}_k\mathbf{h}_k^H\mathbf{W}_k)+\sum\limits_
{\substack{j\neq k}}^K\Tr(\mathbf{h}_k\mathbf{h}_k^H\mathbf{W}_j)\notag\\
&+&\hspace*{-3mm}\Tr(\mathbf{h}_k\mathbf{h}_k^H\mathbf{V})+\sigma_{\mathrm{ant}}^2+\frac{1}{\rho_k}\sigma_{\mathrm{s}}^2\Big]\notag\\
\hspace*{-5mm}&+&\hspace*{-3mm}\sum_{k=1}^K\beta_k\Bigg[\frac{P^{\mathrm{min}}_{k}}
{\eta(1-\rho_k)}-\sigma_{\mathrm{ant}}^2-\Tr\Big(\mathbf{h}_k\mathbf{h}_k^H(\mathbf{V}+\sum_{j=1}^K\mathbf{W}_j)\Big)\Bigg]\notag\\
\hspace*{-5mm}&+&\hspace*{-3mm}\sum_{m=1}^M\nu_m\Bigg[\frac{P^{\min}_{\mathrm{I}_m}}{\eta}-N_{\mathrm{R}}
\sigma_{\mathrm{ant}}^2-\Tr\Big(\mathbf{G}_m\mathbf{G}_m^H(\mathbf{V}+\sum_{k=1}^K\mathbf{W}_k)\Big)\Bigg]\notag\\
\hspace*{-5mm}&+&\hspace*{-3mm}\sum_{m=1}^M\sum_{k=1}^K\Tr\Big\{\mathbf{X}_{m,k}\Big[\mathbf{G}_m^H\mathbf{W}_k\mathbf{G}_m-
(\xi_{\mathrm{eav}_{m,k}}-1)\mathbf{Q}_{m}\Big] \Big\}\notag,
\end{eqnarray}
where $\mathbf{X}_{m,k}$, $\mathbf{Y}$, and $\mathbf{Z}_k$ are the dual variable matrices of constraints $\overline{\mbox{C2}}$ , C6, and C7, respectively. $\alpha_k$, $\beta_k$, and $\nu_m$ are the scalar dual variables of  constraints C1, C3, and C4, respectively.
On the other hand, boundary constraint C5 for $\rho_k$
is  satisfied automatically and the optimal $\rho_k$ will be  illustrated in the later part of this proof.

Then, the dual problem of the SDP relaxed optimization problem in (\ref{eqn:sdp_relaxation}) is given by
\begin{equation}\hspace*{-0mm}\label{eqn:dual}
\underset{\underset{\mathbf{Z}_k,\mathbf{Y},\mathbf{X}_{m,k}\succeq \mathbf{0}}{\nu_m,\beta_k,\alpha_k\ge0}}{\maxo} \,\underset{\underset{\rho_k}{\mathbf{W}_k,\mathbf{V}\in\mathbb{H}^{N_{\mathrm{T}}}}}{\mino} \, {\cal L}\Big(\hspace*{-0.5mm}\mathbf{W}_k,\hspace*{-0.5mm}\mathbf{V},\hspace*{-0.5mm}\rho_k,\hspace*{-0.5mm}\mathbf{Z}_k,\hspace*{-0.5mm}\mathbf{Y},\hspace*{-0.5mm}
\mathbf{X}_{m,k},\hspace*{-0.5mm}\beta_k,\hspace*{-0.5mm}\alpha_k,\nu_m\Big).
\end{equation}

Since the  SDP relaxed optimization problem in (\ref{eqn:sdp_relaxation}) satisfies Slater's constraint qualification and is jointly convex with respect to the optimization variables, strong duality holds and thus solving (\ref{eqn:dual}) is equivalent to solving (\ref{eqn:sdp_relaxation}). We define $\{\mathbf{W}_k^*,\mathbf{V}^*,\rho_k^*\}$ and $\{\mathbf{Z}_k^*,\mathbf{Y}^*,\mathbf{X}_{m,k}^*,\nu_m^*,\beta_k^*,\alpha_k^*\}$ as the optimal primal solution and the optimal dual solution of (\ref{eqn:sdp_relaxation}). Now, we focus on those KKT conditions which are useful in the proof:
\begin{eqnarray}
\hspace*{-3mm}\mathbf{Z}_k^*,\mathbf{X}_{m,k}^*\hspace*{-3mm}&\succeq&\hspace*{-3mm}\mathbf{0},\,\,\alpha_k^*,\,\beta_k^*,\nu_m^*\ge 0,\,\forall k,\,\forall m, \label{eqn:dual_variables}\\
\hspace*{-3mm}\mathbf{Z}_k^*\mathbf{W}_k^*\hspace*{-3mm}&=&\hspace*{-3mm}\mathbf{0},\label{eqn:complementary_cond}\\
\hspace*{-3mm}\mathbf{Z}_k^*\hspace*{-3mm}&=&\hspace*{-3mm}\mathbf{U}_{k}-(\beta_k^*+\frac{\alpha_k^*}{\Gamma_{\mathrm{req},k}})
\mathbf{h}_k\mathbf{h}_k^H,\label{eqn:lagrangian_gradient}\\
\mbox{where }\quad\notag\mathbf{U}_k\hspace*{-3mm}&=&\hspace*{-3mm}\mathbf{I}_{N_{\mathrm{T}}}+\sum_{m=1}^M\mathbf{G}_m(\mathbf{X}_{m,k}^*\hspace*{-1mm}-\hspace*{-1mm}
\nu_m^*\mathbf{I}_{N_{\mathrm{T}}})\mathbf{G}_m^H\\
&+&\hspace*{-3mm}\sum_{j\neq k}^K(\alpha_j^*-\beta_j^*)\mathbf{h}_j\mathbf{h}_j^H,\quad \mbox{and}\\
\rho^*_k\hspace*{-3mm}&=&\hspace*{-3mm}\frac{\sqrt{\alpha_k^*\sigma_{\mathrm{s}}^2\eta}}{\sqrt{\alpha_k^*\sigma_{\mathrm{s}}^2\eta}+\sqrt{\beta_k^* P^{\mathrm{min}}_{k}}} \label{eqn:optimal_rho},\forall k.
\end{eqnarray}
It can be observed from (\ref{eqn:optimal_rho}) that constraint $\mbox{C5: } 0\le \rho_k^*\le 1$ is  automatically satisfied. Besides,  $\alpha^*_k,\beta^*_k>0$  must holds for $\Gamma_{\mathrm{req},k}>0$ and $P_{k}^{\min}>0$. On the other hand, because of the complementary slackness condition on $\mathbf{W}^*_k$ in (\ref{eqn:complementary_cond}), the
columns of $\mathbf{W}_k^*$ are required to lie in the null space of $\mathbf{Z}_k^*$ for $\mathbf{W}_k^*\ne \mathbf{0}$.  In other words, the structure of  $\mathbf{W}^*_k$ depends on the space spanned by $\mathbf{Z}_k^*$. Thus, we focus on the following two cases for revealing the space spanned by $\mathbf{Z}_k^*$. Without loss of generality, we denote $r_k=\Rank(\mathbf{U}_k)$. In the first case, we investigate  the structure of  $\mathbf{W}^*_k$ when $\mathbf{U}_k$ is a full-rank matrix, i.e., $r_k=N_{\mathrm{T}}$. By exploiting (\ref{eqn:lagrangian_gradient}) and a basic  inequality for the rank of matrices, we have
\begin{eqnarray}\label{eqn:rank_inequality}
&&\hspace*{-3mm}\Rank(\mathbf{Z}_k^*)+\Rank((\frac{\alpha_k^*}{\Gamma_{\mathrm{req},k}}+\beta_k^*)\mathbf{h}_k\mathbf{h}_k^H)\ge \Rank(\mathbf{U}_k)\notag\\
\Longleftrightarrow &&\hspace*{-3mm} \Rank(\mathbf{Z}_k^*)\ge N_{\mathrm{T}}-1\quad  \mbox{for} \quad\alpha^*_k,\beta_k^*>0.
\end{eqnarray}
For $\Gamma_{\mathrm{req},k}>0$ and   $\Rank(\mathbf{U}_k)=N_{\mathrm{T}}$, $\Rank(\mathbf{W}_k^*)=1$ and  $\Rank(\mathbf{Z}_k^*)=N_{\mathrm{T}}-1$ must hold simultaneously. Next, we consider the case when $\Rank(\mathbf{U}_k)$ is rank-deficient, i.e., $ r_k<N_{\mathrm{T}}$. Without loss of generality, we define $\nullspace(\mathbf{U}_k)=\mathbf{N}_k$, $\mathbf{N}_k\in\mathbb{C}^{N_{\mathrm{T}}\times (N_{\mathrm{T}}-r_k)}$ such that $\mathbf{U}_k\mathbf{N}_k=\mathbf{0}$ and $\Rank(\mathbf{N}_k)=N_{\mathrm{T}}-r_k$. Let $\boldsymbol{\varrho}_{t_k}\in\mathbb{C}^{N_{\mathrm{T}}\times 1}$, $1\le t_k\le N_{\mathrm{T}}-r_k$, denote the $t_k$-th column vector of $\mathbf{N}_k$.  Then, by exploiting (\ref{eqn:lagrangian_gradient}),  we have the following equality:
\begin{eqnarray}\label{eqn:pre-post}
\boldsymbol{\varrho}_{t_k}^H\mathbf{Z}_k^*\boldsymbol{\varrho}_{t_k}=- (\frac{\alpha^*_k}{\Gamma_{\mathrm{req},k}}+\beta^*_k)\boldsymbol{\varrho}_{t_k}^H\mathbf{h}_k\mathbf{h}_k^H\boldsymbol{\varrho}_{t_k}.
\end{eqnarray}
 Combining  $\mathbf{Z}^*_k\succeq \mathbf{0}$ and $\frac{\alpha^*_k}{\Gamma_{\mathrm{req},k}}+\beta^*_k>0$,   $(\frac{\alpha^*_k}{\Gamma_{\mathrm{req},k}}+\beta^*_k)\boldsymbol{\varrho}_{t_k}^H\mathbf{h}_k\mathbf{h}_k^H\boldsymbol{\varrho}_{t_k}={0},\forall t_k\in\{1,\ldots,N_{\mathrm{T}}-r_k\}$, holds in (\ref{eqn:pre-post}). In other words,
\begin{eqnarray}\label{eqn:rank-nullspace}
\mathbf{Z}^*_k\mathbf{N}_k=\mathbf{0} \quad \mbox{ and }\quad \mathbf{h}_k\mathbf{h}_k^H\mathbf{N}_k=\mathbf{0}
\end{eqnarray}
hold and the columns of $\mathbf{N}_k$  lie in the null spaces of $\mathbf{h}_k\mathbf{h}_k^H$ and $ \mathbf{Z}^*_k$ simultaneously. Furthermore, $\Rank\Big(\nullspace(\mathbf{Z}^*) \Big)\ge N_{\mathrm{T}}-r_k$ holds for satisfying  $\mathbf{Z}_k^*\mathbf{N}_k=\mathbf{0}$. On the other hand, from (\ref{eqn:rank_inequality}) and $\Rank(\mathbf{U}_k)=r_k$, we obtain
\begin{eqnarray}\label{eqn:temp}
 \Rank(\mathbf{Z}_k^*)\ge r_k-1.
\end{eqnarray} Then, by utilizing  (\ref{eqn:rank-nullspace})  and (\ref{eqn:temp}),  $\Rank\Big(\nullspace(\mathbf{Z}^*_k) \Big)$ is bounded between
\begin{eqnarray}
&&N_{\mathrm{T}}- r_k+1   \ge  \Rank\Big(\nullspace(\mathbf{Z}^*_k) \Big)\ge N_{\mathrm{T}}-r_k.
\end{eqnarray}
As a result, either  $\Rank\Big(\nullspace(\mathbf{Z}^*_k) \Big)=N_{\mathrm{T}}-r_k$ or $\Rank\Big(\nullspace(\mathbf{Z}^*_k)\Big)=N_{\mathrm{T}}-r_k+1 $ holds for the optimal solution. Suppose $\Rank\Big(\nullspace(\mathbf{Z}^*_k) \Big)=N_{\mathrm{T}}-r_k$ and thus  $\nullspace(\mathbf{Z}^*_k)= \mathbf{N}_k$.  Then, we can express $\mathbf{W}^*_k$ as $\mathbf{W}^*_k=\sum_{t_k=1}^{N_{\mathrm{T}}-r_k} \gamma_{t_k}\boldsymbol{\varrho}_{t_k} \boldsymbol{\varrho}_{t_k}^H$  for some positive constants $\gamma_{t_k}\ge 0,\forall t_k\in\{1,\ldots,N_{\mathrm{T}}-r_k\} $. Yet,  due to (\ref{eqn:rank-nullspace}),
\begin{eqnarray}\Tr\big(\mathbf{h}_k\mathbf{h}^H_k\mathbf{W}^*_k\big)=\sum_{t_k=1}^{N_{\mathrm{T}}-r_k}\gamma_{t_k}
\Tr\big(\boldsymbol{\varrho}_{t_k}^H\mathbf{h}_k\mathbf{h}^H_k\boldsymbol{\varrho}_{t_k} \big)=0
\end{eqnarray}
holds which cannot satisfy constraint C1 for $\Gamma_{\mathrm{req}_k}>0$. Thus, $\Rank\Big(\nullspace(\mathbf{Z}^*_k)\Big)=N_{\mathrm{T}}-r_k+1$ has to hold for the optimal  $\mathbf{W}^*_k$. Besides,  there exists one subspace spanned by a unit norm vector  $\mathbf{u}_k\in \mathbb{C}^{N_{\mathrm{T}}\times 1}$  such that $\mathbf{Z}^*_k\mathbf{u}_k=\mathbf{0}$ and $\mathbf{N}_k^H\mathbf{u}_k=\mathbf{0}$. Therefore,  the orthonormal  null space of $\mathbf{Z}^*_k$ can be presented as
\begin{eqnarray}
\nullspace(\mathbf{Z}^*_k)=\Big\{\mathbf{N}_k\cup\mathbf{u}_k\Big\}.
\end{eqnarray}
In summary, without loss of generality, we can express the optimal solution of  $\mathbf{W}^*_k$  as
\begin{eqnarray}\label{eqn:general_structure}
\mathbf{W}^*_k=\sum_{t_k=1}^{N_{\mathrm{T}}-r_k} \gamma_{t_k}\boldsymbol{\varrho}_{t_k} \boldsymbol{\varrho}_{t_k}^H  + f_k\mathbf{u}_k\mathbf{u}_k^H,
\end{eqnarray}
 where  $f_k>0 $ is some positive scaling constant.

In the second part of the proof, for $\Rank(\mathbf{W}^*_k)>1$,  we reconstruct another solution of the relaxed version of problem (\ref{eqn:sdp_relaxation}),  $\{\mathbf{\widetilde  W}_k ,\mathbf{\widetilde  V},\widetilde \rho_k\}$, based on  (\ref{eqn:general_structure}).

Let the constructed solution set be given by
\begin{eqnarray}\label{eqn:rank-one-structure}\mathbf{\widetilde W}_k&=&f_k\mathbf{u}_k\mathbf{u}_k^H=\mathbf{W}^*_k-\sum_{t_k=1}^{N_{\mathrm{T}}-r_k} \gamma_{t_k} \boldsymbol{\varrho}_{t_k} \boldsymbol{\varrho}_{t_k} ^H, \\
 \mathbf{\widetilde V}&=&\mathbf{ V^*}+\sum_{t_k=1}^{N_{\mathrm{T}}-r_k} \gamma_{t_k} \boldsymbol{\varrho}_{t_k} \boldsymbol{\varrho}_{t_k}^H, \quad\widetilde \rho_k=\rho^*_k \label{eqn:rank-one-structure2}.
\end{eqnarray}
It can be easily verified  that   $\{\mathbf{\widetilde W}_k,\mathbf{\widetilde V},  \widetilde \rho_k\}$ not only satisfies the constraints in (\ref{eqn:sdp_relaxation}), but also achieves the same optimal objective value as $\{\mathbf{ W}_k,\mathbf{ V},  \rho_k\}$ with $\Rank(\mathbf{\widetilde W}_k)=1,\forall k$. The actual values of  $\{\mathbf{\widetilde W}_k,\mathbf{\widetilde V},  \widetilde \rho_k\}$ can be obtained by substituting  (\ref{eqn:rank-one-structure}) and (\ref{eqn:rank-one-structure2}) into  (\ref{eqn:sdp_relaxation}) and solving the resulting  convex optimization problem for  $f_k$ and $\gamma_{t_k}$.

\bibliographystyle{IEEEtran}
\bibliography{OFDMA-AF}

\end{document}